\documentclass{elsart}
\usepackage{graphicx}
\newcommand{\mb}[1]{ { \mbox{\boldmath{$#1$}}}  }
\newcommand{\mbs}[1]{ {\scriptsize \mbox{\boldmath{$#1$}}}  }
\newcommand{\sub}[2]{ \mbox{$#1$}_{\mbox{\scriptsize\boldmath{$#2$}}}}

\begin{document}
\begin{frontmatter}
\title{
Spatial fluctuations of the pairing  potential  in
disordered superconductors}
\author{Grzegorz Litak\thanksref{E-mail}}
\address{
Department of Mechanics, Technical University of
Lublin Nadbystrzycka 36, PL--20-618  Lublin, Poland.}

\thanks[E-mail]{Fax: +48-815250808; E-mail:
litak@archimedes.pol.lublin.pl}

\begin{abstract}
We study the effect of site  diagonal, non-magnetic, disorder on the
a pairing amplitude in an extended  Hubbard model with the intersite
attraction.
Analyzing  fluctuations of a pairing potential we discuss
 the instability of
mixed solutions, '$s+d$' and '$s+{\rm i} d$', in presence of disorder.
The influence of disorder on extended $s$-- and $d$--wave superconductors
appear to be comparable but in certain regions of the
phase diagram, even weak disorder can change the symmetry of the
order parameter. 
\end{abstract}
\begin{keyword}
\sep exotic pairing \sep doping  \sep disorder  
\PACS{74.62.Dh, 74.20-z}
\end{keyword}

\end{frontmatter}

\section{Introduction}

 The study of disordered
 superconductors with exotic  pairing is of
general
interest. 
$D$--wave pairing
plays an important role in the superconductivity of the cuprates
\cite{Ann96} while $p$--wave in strontium ruthenate (Sr$_2$RuO$_4$)
\cite{Mae01}.
Depending on symmetry of order parameter, the effect of disorder
is dramatically
different on the superconducting states which are described as having
isotropic, $s$--wave, or anisotropic, extended $s$--,  $d$-- and
$p$--wave, 
order parameter
symmetry \cite{And59,Abr59,Mak69,Lus73,Gor83,Bor94,Gyo97,Lit98b,Lit98a,Mar99,Lit00}. 
The works on disorder effect on superconductors are stimulated by
the recent experiments \cite{Bon93,Bon94,Kar00,Ber96,Tal97,Mac98}.

In this paper we analyze  an extended  Hubbard model with intersite
attraction,
whose phase diagram includes both $s$-- and $d$--wave regions and
introduce
disorder into the problem by allowing the site energies, $\varepsilon_i$,
to be independent random variables. 

 In  the  case of anisotropic $s$-- and $d$--wave 
pairing there is no strict equivalent of the Anderson Theorem 
\cite{And59,Mak69,Gyo97} which
governs the response of isotropic $s$--wave 
superconductors to randomness in the
crystal potential. However,  a weaker
statement, which
may
be
similarly
useful, can be formulated \cite{Gor83,Lit98a,Mar99}.
Here we  will calculate 
the ensemble average $<\Delta_{ij} \Delta_{il}^*>$ where
$\Delta_{ij}$ is the pairing amplitude in the cases where the 
sites, $i,j$ and $l$, are nearest neighbours to lowest order in the 
fluctuations of the site
energies $\delta \varepsilon_i= \varepsilon_i -\varepsilon_0$ about
their mean $\varepsilon_0$. Our results 
imply that the disorder does not induce large fluctuations in the
pairing potentials and hence for systems of large coherence length
$\xi$ the amplitude of pairing potential $|\Delta(ij)|$ may rather to be
the same for all
bonds in $x$  and $y$ directions \cite{Gyo97,Lit98a,Mar99}.
Eventually, we compare the corresponding standard square deviation
of  fluctuating potentials $\sigma (\delta \Delta_{ij})$ and $\sigma (\delta
\varepsilon_{i})$ (Fig. 1). Their ratio $\Gamma= ( \sigma (\delta
\Delta_{ij})/\sigma 
(\delta 
\varepsilon_{i}) )^2$ will be a criterion of pairing potential
fluctuations 
leading finally to
the destruction of superconducting phase.  
We shall be also interested what is  the effect of disorder on anisotropic
$s$-- and $d$--
wave superconductors, where they coexist,  and if the disorder favour any
of
particular solution. 
\begin{figure}[htb]
\hspace{2cm}
\resizebox{0.3\textwidth}{!}{%
  \includegraphics{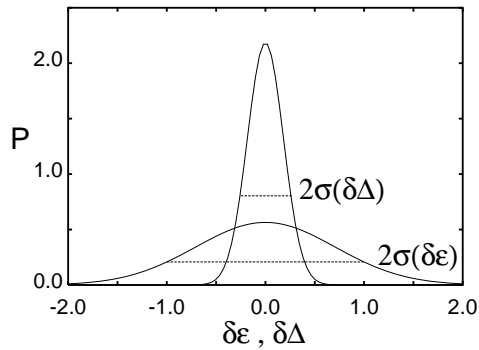}}   
\caption{The examples of Gaussian distribution for $\delta \varepsilon_i$ with
a mean square deviation $\sigma (\delta \varepsilon_i)= \sqrt{<
\varepsilon_i^2>}$
and the expected distribution of  a pairing potential $\delta \Delta_i$
with
corresponding $\sigma (\delta \Delta_{ij})= \sqrt{<
|\Delta_{ij}|^2>}$.  Fluctuating potentials $\delta \varepsilon_i$ and
$\delta
\Delta_{ij}$ are in the units of $\sqrt{< \varepsilon_i^2>}$}
\end{figure}

The paper is organized as follows. In Sec. 2 we present the model in
the
clean limit, introduce the Hamiltonian and approximations used in paper.
Here we also discuss briefly
the phase diagram of the clean system. In Sec. 3 we include 
disorder
induced spatial fluctuations of
a pairing
potential. Here we discuss the formalism used in our analysis and show
the results
on the disordered effect on
anisotropic superconductors. Finally we investigate the  stability of
various solutions in
presence of disorder.
Section 4 contains conclusions and remarks.

\section{ The model in clean limit.
}

Here we use a single band, extended, Hubbard model with an intersite
attraction. It is defined by the
Hamiltonian \cite{Mic90}:
\begin{equation}
H=  \sum_{ij \sigma} (\varepsilon_i \delta_{ij} +t_{ij}
)c_{i\sigma}^{+}c_{j\sigma} + \sum_{i j \sigma \sigma'} 
\frac{W_{ij}}{2} n_{i \sigma} n_{j \sigma'}
- \mu \sum_{i \sigma}
c_{i\sigma}^+c_{i\sigma},
\label{eq1}
 \end{equation}
 
\noindent where $c_{i\sigma}^+$ and $c_{i \sigma}$ are usual, fermionic
operators    which     create     and 
annihilate, respectively, an electron 
with spin
$\sigma$  at the
lattice
site labeled by $i$, $t_{ij}$ is a electron hopping integral, $n_{i \sigma}$
is the operator of particle number  of spin $\sigma$ at site $i$,
$\varepsilon_i$ is the 
site
energy, varying from site to site 
in random fashion,  at the site $i$ with mean value
$\varepsilon_0=<\varepsilon_i>=0$ and $W_{ij}$ is the 
interaction potential
of two electrons with opposite spins on neighbour sites $i,j$. Finally, $\mu$ denotes the
chemical
 potential.

Hartree--Fock--Gorkov equation of motion for Gorkov Greens
$2 \times
2$ functions ${\mb G}(i,j;\omega)$ yields:

\begin{eqnarray}
\sum_l&~&
\left[ \begin{array}{cc} (\omega-\varepsilon_l  +\mu ) \delta_{il}-t_{il} - W_{il} (
<c_{i \downarrow}^+ c_{l \downarrow}> - n_l)
~~~~~W_{il}<c_{i \downarrow} 
c_{l \uparrow}> \\
W_{il}<c_{i \uparrow} 
c_{l \downarrow}>~~~~~ (\omega+\varepsilon_l   -\mu )
\delta_{il}+t_{il} + 
W_{il}( <c_{i \uparrow}^+ c_{l \uparrow}> - n_l) \end{array}
\right] \nonumber \\ &\times&
{\mb G}(l,j; \omega)={\mb 1} \delta_{ij},
\label{eq2}
\end{eqnarray}

\noindent where we assumed a paramagnetic state $n_{l \uparrow}=n_{l \downarrow}= \frac{n_l}{2}$.

The charge $n_i$ on the site $i$ and pairing potential 
$\Delta_{il} =W_{il}<c_{i \downarrow} 
c_{l \uparrow}> $ for neighbour sites $i,l$ can 
be expressed by following equations:

\begin{eqnarray}
n_i &=& - \frac{2}{\pi}
 \int_{-\infty}^{\infty} {\rm d} \omega~
{\rm Im}~ G_{11} (i,i;\omega ) \frac{1}{{\rm e}^{\beta \omega} +   
1}, \nonumber \\
\Delta_{il} &=&  - \frac{W_{il}}{\pi} \int_{-\infty}^{\infty}
  {\rm d} \omega~
{\rm Im}~ G_{12} (i,l;\omega) \frac{1}{{\rm e}^{\beta \omega} +
1},
\label{eq3}
\end{eqnarray}
where $\beta=\frac{1}{k_BT}$ and $k_B$ is the Boltzmann constant.

For a pure system we take the lattice Fourier transform of Eq.
\ref{eq2} with
$\varepsilon_i=0$ and  find:

\begin{equation}
\left[ \begin{array}{cc} \omega-\sub{\epsilon}{k}'  +\mu'    & \sub{\Delta}{k}
 \\
\sub{\Delta}{k}^* &   \omega
+\sub{\epsilon}{ k}' -\mu'   \end{array}      \right] {\mb G}^0(\mb
k; \omega)={\mb 1}
\label{eq4},
\end{equation}
            \noindent where $\mu'$ is shifted
chemical potential $ \mu'=\mu+ Wn$, 
and the Hartree--Fock kinetic energy $\sub{\epsilon}{k}$ is given by
\begin{equation}
\sub{\epsilon}{k}'=\sub{\epsilon}{k}-
\frac{W}{N}\sum_{\mb k} \sub{n}{k} \sub{\gamma}{k},
\label{eq5}
\end{equation} 
where 
\begin{equation}
\label{eq6}
\sub{n}{k}= < \sub{c}{k} \sub{c}{k}^+> =\sum_{ij}  <
c_i^+
c_j>{\rm
e}^{{\rm i} (\mbs R_j-\mbs R_i) \mbs
k}. \sub{~}{~}^{~}
\end{equation}
Following Ref. \cite{Mic90}  we will neglect the Fock term 
$ \frac{W}{N} \sum_{\mbs k} \sub{n}{k}
\sub{\gamma}{k}$ (Eq. \ref{eq5}). Thus, for two dimensional lattice and
the electron hopping defined
between  nearest
neighbour sites only,  we get
\begin{equation}
\sub{\epsilon}{k}' \approx \sub{\epsilon}{k} 
 = \sum_i
t_{ij}
{\rm
e}^{{\rm i} (\mbs R_j-\mbs R_i) \mbs
k}=-  t \sub{\gamma}{k},
\label{eq7}
\end{equation}
 where 
\begin{equation}
\sub{\gamma}{k}= 2 ({\cos} k_x +{\cos} k_y).
\label{eq8}
\end{equation}

\noindent As can be readily shown 
the order parameter $\sub{\Delta}{k}$  satisfies the following gap
 equation:

\begin{equation}
\label{eq9}
\sub{\Delta}{k} = 
  \frac{1}{N} \sum_{\mbs q} \frac{\sub{W}{ k -q} \sub{\Delta}{q}}
{ 2 \sub{E}{q}} {\rm tanh} \left( \frac{\beta \omega}{2} \right),
\end{equation}
where 
\begin{eqnarray}
\sub{W}{k-q} &=& |W| \sub{\gamma}{k-q}
\label{eq10}  \\ 
 &=& |W| \left( \frac{\sub{\gamma}{k} \sub{\gamma}{q} +\sub{\eta}{k}
\sub{\eta}{q}}{4} + 2 {\rm sin}(k_x) {\rm sin}(q_x) + 2 {\rm sin}(k_y) {\rm sin}(q_y)
\right)
\nonumber \end{eqnarray}
and 
\begin{equation}
\label{eq11}
\sub{\eta}{k} = 2 ({\rm cos} k_x -{\rm cos} k_y ),  
\end{equation}
while $\sub{E}{q}$ denotes quasi-particle energy: 
\begin{equation}
\sub{E}{q} = \sqrt{\sub{\tilde{\epsilon}}{q}^2 -\sub{\Delta}{q}^2}~~,~~~~~~ 
\sub{\tilde{\epsilon}}{q}= \sub{\epsilon}{q} -\mu'.
\label{eq12} 
\end{equation}

Generally, the singlet type of solution (Eq. \ref{eq9}) can be written as 
\begin{equation}
\sub{\Delta}{k} = \Delta^{s} \sub{\gamma}{k} + \Delta^{d} \sub{\eta}{k} 
\label{eq13}
\end{equation}
for the real type solution if $s$ and $d$ parts of $\Delta(\mb k)$
have the same phase and
\begin{equation}
\sub{\Delta}{k} =
\Delta^{s} \sub{\gamma}{k} + {\rm i} \Delta^{d} 
\sub{\eta}{k}
\label{eq14} 
\end{equation}
for the complex solution.
Pairing amplitudes: $\Delta^{s}$,
$\Delta^{d}$ are defined as real numbers, corresponding to  $s$-- and
$d$--wave
components
respectively.
 
The pairing parameters $\Delta^{s}$,
$\Delta^{d}$  (Eqs. \ref{eq13}, \ref{eq14}) may be
determined from Eq. (\ref{eq9}). Namely:

\begin{eqnarray}
 \Delta^{s} &=& -\frac{W}{N}\sum_{\mbs q} \frac{\sub{\gamma}{q}}{ 8
\sub{E}{q}}
~~{\rm tanh} \left( \frac{\beta \sub{E}{q}}{2} \right) \sub{\Delta}{q},
 \nonumber \\
\Delta^{d} \alpha &=& -\frac{W}{N}\sum_{\mbs q} \frac{\sub{\eta}{q}}{ 8
\sub{E}{q}}
~~{\rm tanh} \left( \frac{\beta \sub{E}{q}}{2} \right) \sub{\Delta}{q},
\label{eq15}
\end{eqnarray}
where
and $\alpha = 1$ for  a real solution ($s+ d$)
while $\alpha = {\rm i}$ for
a complex one ($s+ {\rm i}d$).

The above set of equations for pairing potentials (\ref{eq15}) should be
completed by
the equation for the
chemical
potential $\mu$ (Eq. \ref{eq3}). It can be written as:
\begin{equation}
n-1= -\frac{2}{N} \sum_{\mbs q} \frac{\sub{\tilde{\epsilon}}{q}}{ 2
\sub{E}{q}} 
~~{\rm tanh} \left( \frac{\beta \sub{E}{q}}{2} \right).
\label{eq16} 
\end{equation}

The gap equations (Eqs. \ref{eq15}) 
can be used  to calculate the
superconducting critical temperature $T_C$ after the linearization. 
The two separate equations
read:

\begin{eqnarray}
1 &=& -\frac{W}{N}\sum_{\mbs q} \frac{{\sub{\gamma}{q}}^2}{ 8
\sub{\tilde \epsilon }{q}}
~~{\rm tanh} \left( \frac{\beta_C \sub{\tilde \epsilon}{q}}{2} \right),
 \nonumber \\
1 &=& -\frac{W}{N}\sum_{\mbs q} \frac{{\sub{\eta}{q}}^2}{ 8
\sub{\tilde \epsilon}{q}}
~~{\rm tanh} \left( \frac{\beta_C \sub{\tilde \epsilon}{q}}{2} \right),
\label{eq17}  
\end{eqnarray}
where $\beta_C=\frac{1}{k_BT_C}$.
 
In the Fig. 2  we present $T_C$ results 
 for two dimensional lattice versus  band filling $n$.
 The interaction parameter $W$ was chosen as  $W/D=-0.3$, $-0.5$, $-0.7$,
where $D=8t$
denotes a bandwidth. 
\begin{figure}[htb]
\hspace{4cm}
\resizebox{0.3\textwidth}{!}{%
  \includegraphics{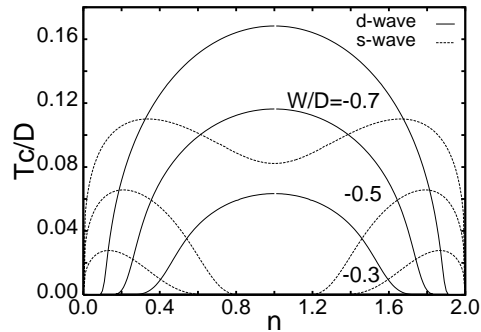}}
\caption{ The superconducting critical temperature
$T_C$ versus band filling $n$ for $s$- and $d$-wave pairing (dashed
and full lines respectively) for three
values of intersite attraction $|W|/D=0.3$, $0.5$, $0.7$}
\end{figure}
Because of  the assumed form of a dispersion relation
$\sub{\varepsilon}{k}$ (Eq. \ref{eq7})
 the model (Eq. \ref{eq1}) possesses the particle--hole symmetry,
therefore
 $T_C$ for  band--filling $n$ and $2-n$  is the same. 
The results show
 that  $s$--wave type of superconductivity exists for 
low electron or hole concentration 
 while $d$--wave type exists close to half--band filling \cite{Mic90}. 
 For a large enough  electron interaction parameter $W/D > 0.2 $ 
the curves $T_C=T_C(n)$
for 
 $s$- and $d$-wave pairing cross and  regions of $s$- and $d$- wave
  superconductivity are not separated.
The pairing parameters, 
$\Delta^s$ and $\Delta^d$ at zero temperature ($T=0$ K) should be
 calculated from the full set of equations Eqs.
(\ref{eq15}). 
For the real type solution (Eq. \ref{eq13}) they are plotted 
for the same three
values of the electron
attraction $W/D=-0.3$,
$-0.5$, $-0.7$ in Fig. 3a. 
Note that in all cases there are small but visible regions of
mixed
'$s+d$'
solutions.
In Fig. 3b we have plotted the same for the complex solution (Eq.
\ref{eq14}) of mixed
'$s+{\rm i} d$' pairing.
Clearly, 
the regions  with possible mixed solution in case of  '$s+{\rm i}
d$' pairing are much larger than for   '$s+d$' mixture.
We will consider these two solutions   more carefully.

\begin{figure}[htb]
\hspace{-1cm}
\resizebox{0.3\textwidth}{!}{%
  \includegraphics{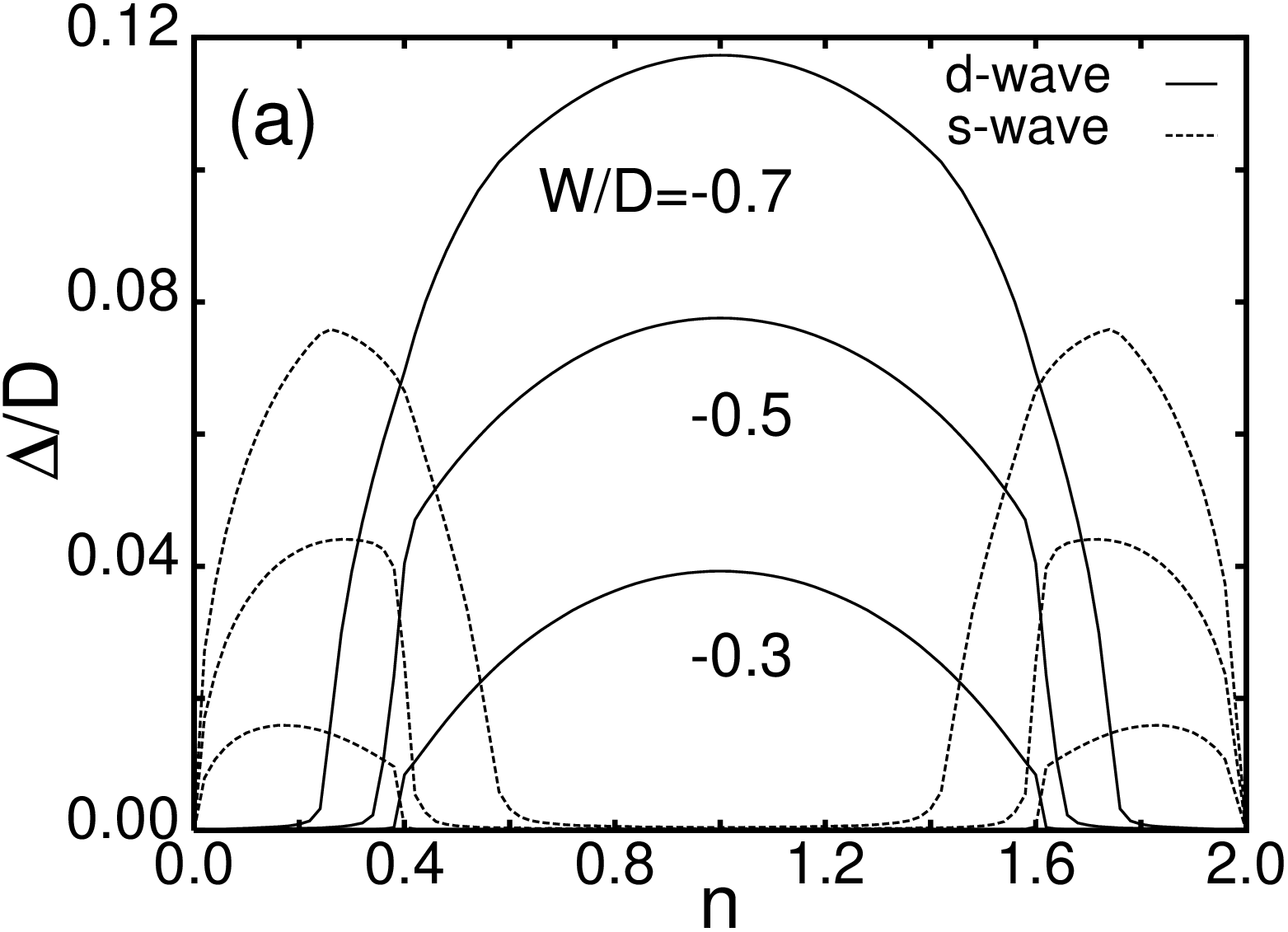}}
\hspace{2cm}
\resizebox{0.3\textwidth}{!}{%
  \includegraphics{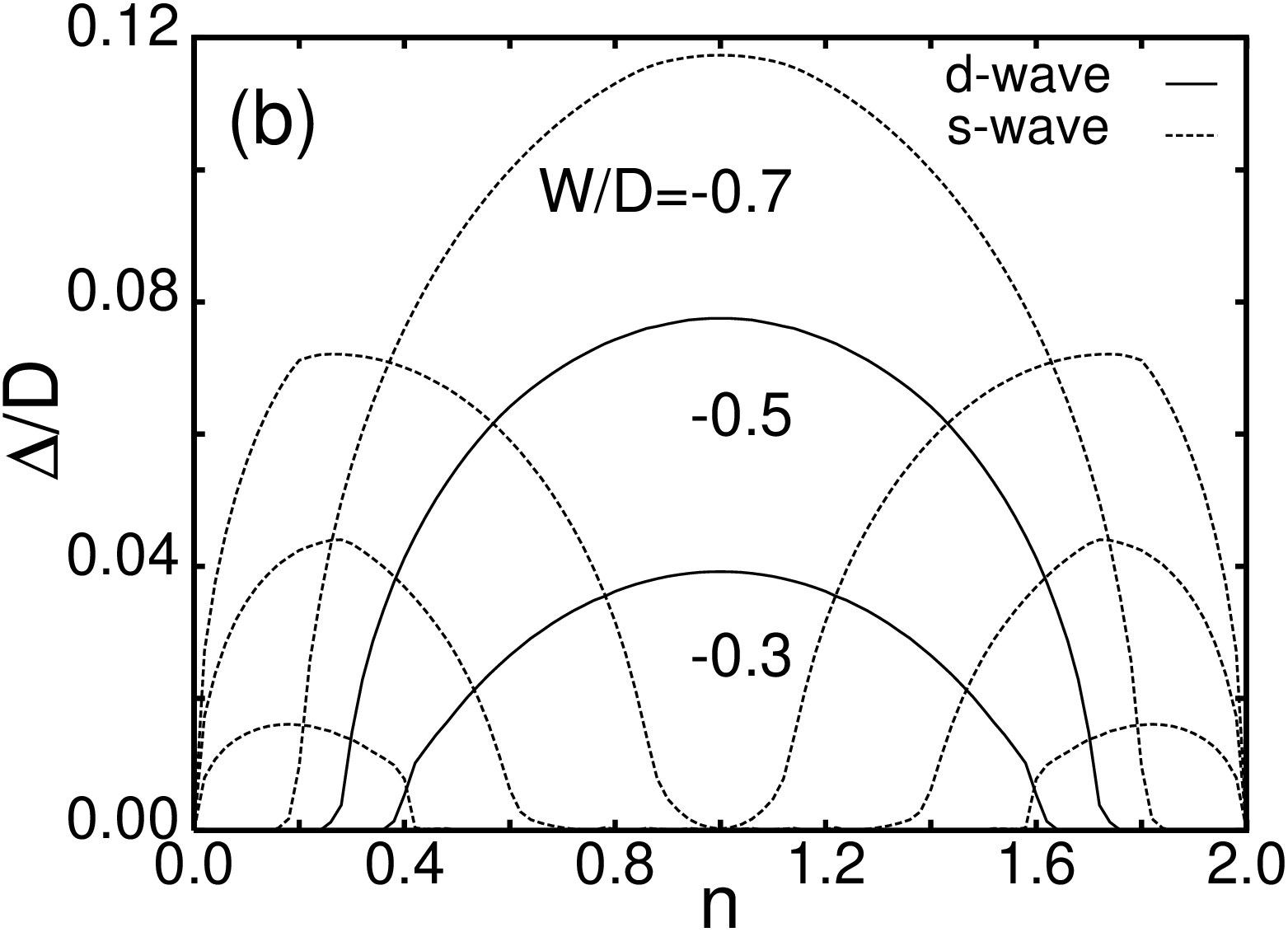}}
\caption{Amplitude of the pairing potential
$\Delta$ versus band filling $n$ for $s$- and $d$- wave pairing (dashed
and full lines respectively) for three
values of intersite attraction $|W|/D=0.3$, $0.5$, $0.7$.
Note the difference around $n=0.4$, figure (a) corresponds to the real
solution '$s+d$' while
(b) to the complex one  '$s+{\rm i} d$' respectively.}
\end{figure}
In the case of mixed solution, the
order
parameter (complex or real) in the lattice real space $\Delta_{ij}$ is
defined as the sum of 
$\Delta^s_{ij}$ and $\Delta^d_{ij}$.
Namely:
\begin{equation}
\Delta_{ij} = \Delta_{ij}^s + \alpha (-1)^{l-j}
\Delta_{il}^d,
\label{eq18}
\end{equation}
where $l$ denotes the neighbour site ($l=j$, $j+1$, $j+2$, $j+3$ Fig. 4).
$\alpha$ is equal to $1$ and ${\rm i}$ for a real and  complex solution,
respectively.

\begin{figure}[htb]
\centering
\vspace{0.5cm}
\resizebox{0.2\textwidth}{!}{%
  \includegraphics{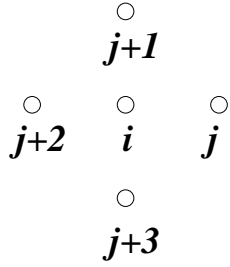}}
\vspace{0.5cm}
\caption{Neighbour sites  on two a dimensional square lattice $i$, $j$
and $l=j+1$, $j+1$, $j+2$,  
$j+3$.}
\end{figure}

From the above equation (Eq. \ref{eq18}) one can determine (Fig. 4) the
$s$ and $d$
components:
\begin{eqnarray}
\Delta_{ij}^s &=& \frac{1}{2} (\Delta_{ij} +\Delta_{ij+1}).
\nonumber \\
\Delta_{ij}^d &=& \frac{1}{2\alpha} (\Delta_{ij} -\Delta_{ij+1})
\label{eq19}
\end{eqnarray}

\begin{figure}[htb]
\hspace{3cm}
\resizebox{0.3\textwidth}{!}{%
  \includegraphics{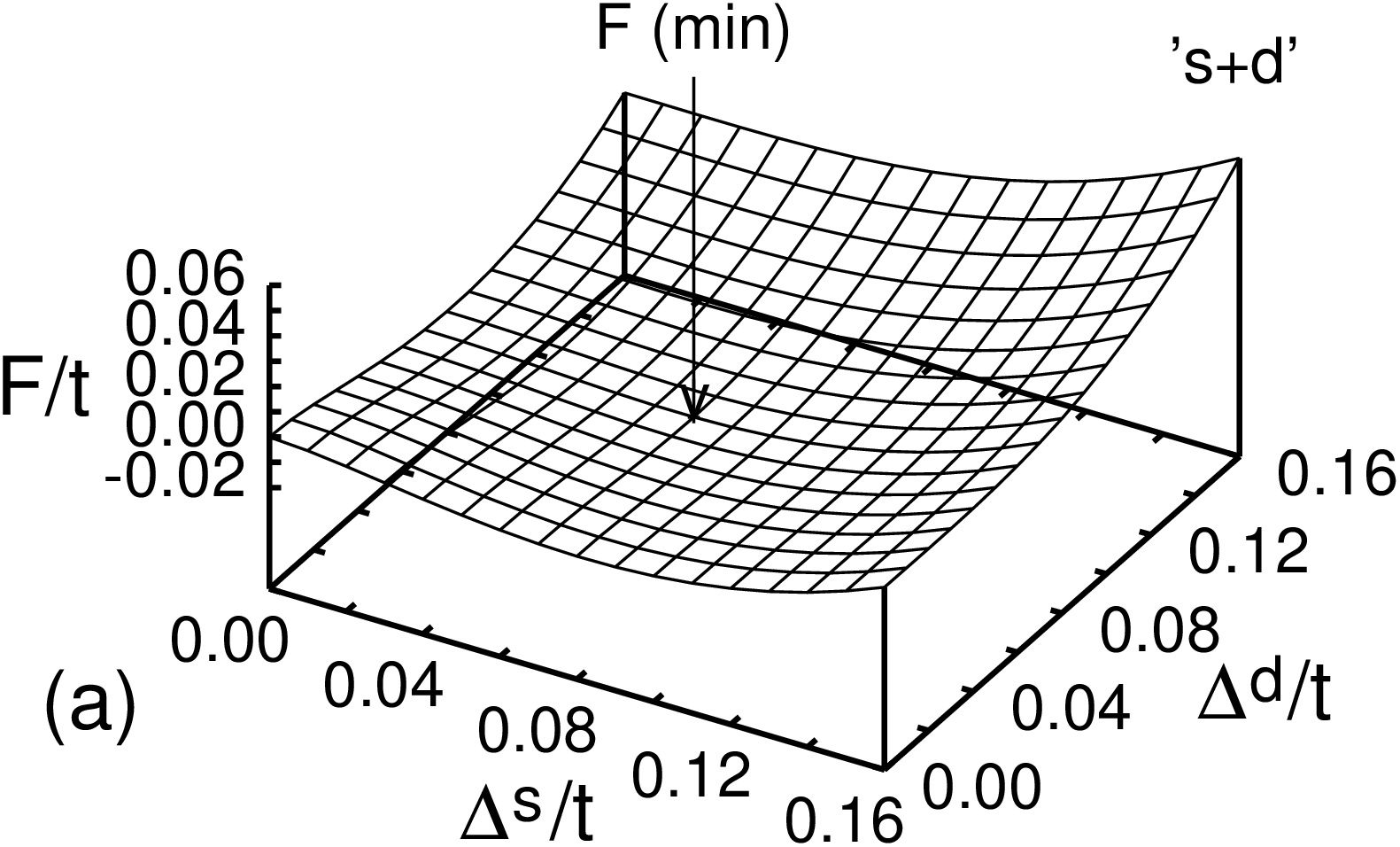}}  

\vspace{-1cm}
\hspace{3cm}
\resizebox{0.3\textwidth}{!}{%
  \includegraphics{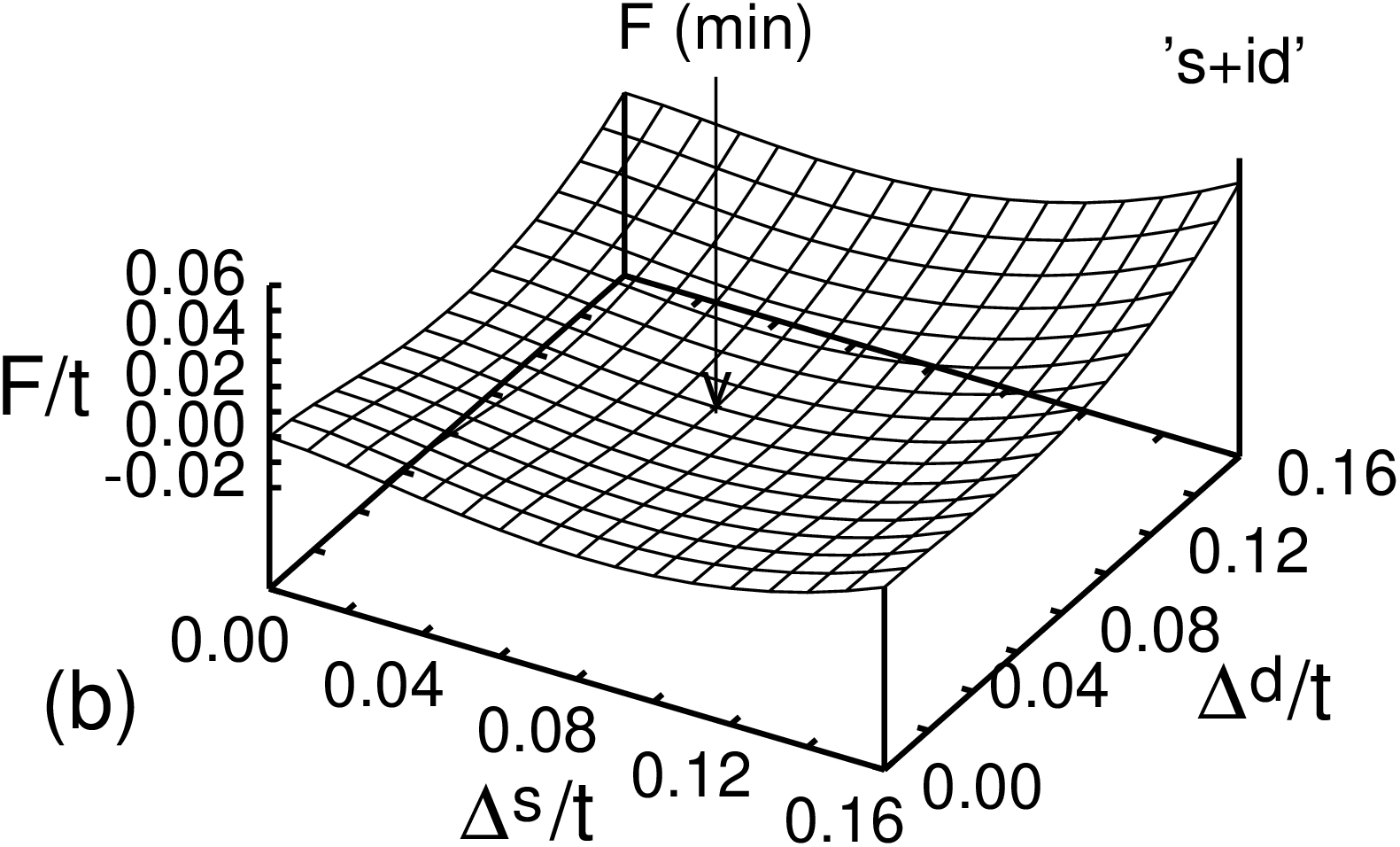}}

\vspace{-1cm}
\hspace{3cm}
\resizebox{0.3\textwidth}{!}{%
  \includegraphics{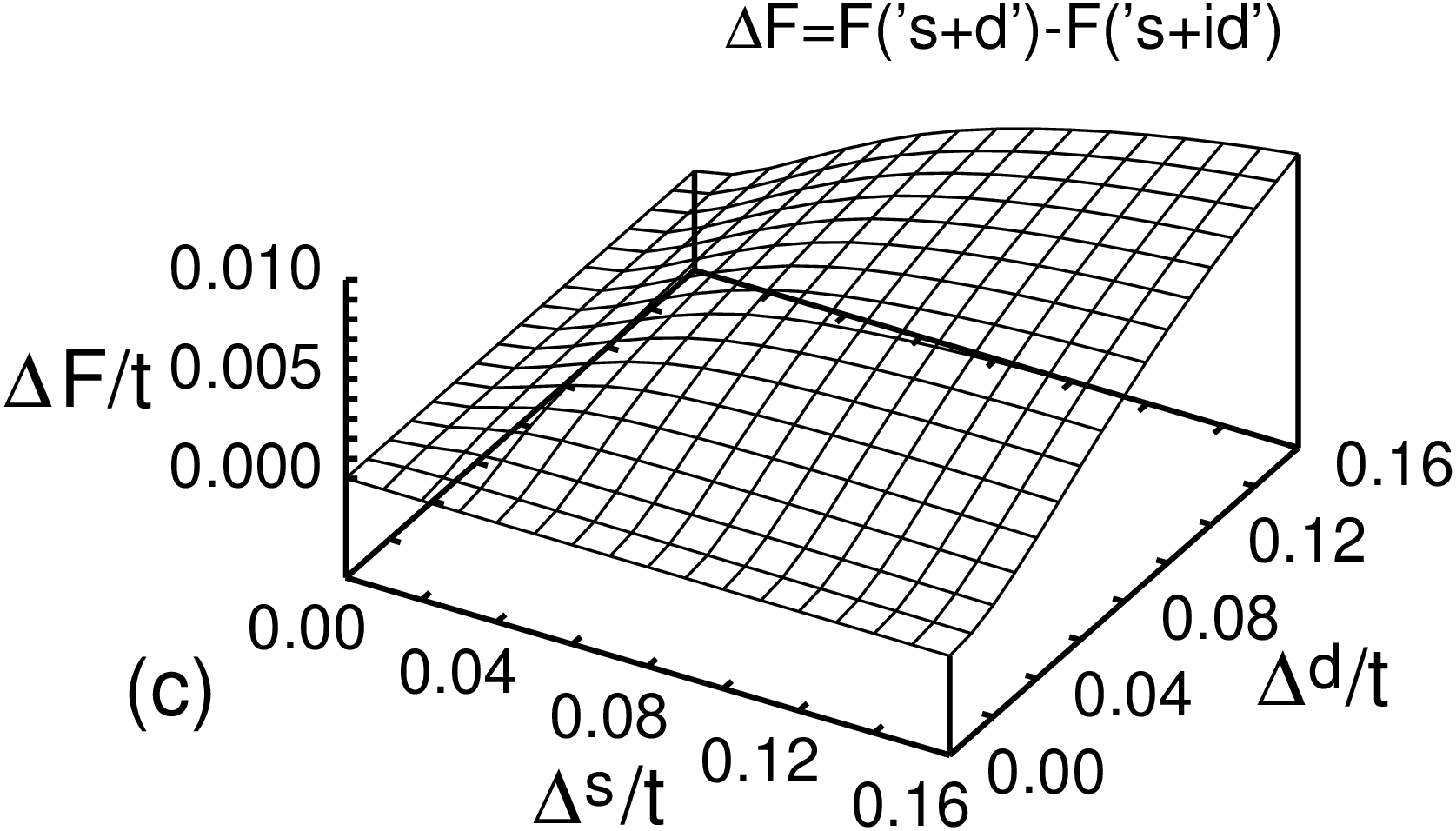}}
\caption{Free  energies $F$ of mixed solutions: (a) the real combination
'$s+d$' and
(b) the complex one '$s+{\rm i} d$' for the same band filling $n=0.4$ and
the attraction parameter $|W|/D=0.7$.
 Additional arrows show the minima of free energies $F$.}
\end{figure}

The free energy $F$ for a finite temperature $T$ can be calculated from
the
following formula \cite{Mic90}:
\begin{equation}
\label{eq20}
F=\frac{1}{N}\sum_{\mbs k} \left[ -(n-1) \sub{\epsilon}{k} -2k_BT~ {\rm ln} \left( 2
\cosh
\frac{ \sub{E}{k} }{2k_BT} \right) -\frac{|\sub{\Delta}{k}(T)|^2
}{W} \right]
\end{equation}
the corresponding derivatives determine 
the integral equations (Eqs. \ref{eq14} and \ref{eq15}):
\begin{equation}
\label{eq21}
\frac{\partial F}{\partial \mu} =0, ~~~~~~~~~~~~\frac{\partial F}{\partial 
\Delta^{
\psi}} =0,~~~~~\psi=s {~~~\rm or~~~} d.
\end{equation}

In Fig. 4  we have plotted  the free energies $F$ of both mixed solutions:
the real
combination
'$s+d$' (Fig. 5a) as well as
the complex one '$s+{\rm i} d$' (Fig. 5b) versus $s$-- and $d$--wave
amplitudes $\Delta^s$
and $\Delta^d$. The band filling $n$ was chosen to be 
$n=0.4$ and
the attraction parameter $|W|/D=0.7$ is 
large enough to produce the mixed solutions
(Figs. 2-3).
Additional arrows in the plots show the minima of free energies $F$.
Clearly 
they show
the mixed type of solutions.
To check which solution is more favorable  we show also the  
difference between free
energies for two
superconducting mixed states    $\Delta F= F('s+ d') -F('s
+ {\rm i} d')$ versus $s$-- and $d$--wave amplitudes (Fig. 5c).
 One can see that in the region of
coexistence $\Delta F$ is positive. Thus, the solution with 
$s+ {\rm i} d$ symmetry is
favored by smaller free energy
$F$.

\section{ Disorder induced Fluctuations of pairing potential}

To go further we apply the same strategy as in Refs. 
\cite{Gyo97,Lit98b,Mor01,Lit96} we
treat random site energies $\varepsilon_i$ as
perturbations and we proceed by solving the Dyson equation for
the Gorkov matrix of the pure superconductor evaluated at a frequency
$\omega$

\begin{equation}
\label{eq22}
{\mb G}(i,j;\omega) = {\mb G}^0(i,j;\omega) + \sum_l
{\mb G}^0(i,j;\omega) {\mb V}_l {\mb G}(l,j;\omega),
\end{equation}
where $\mb V_l$ is the impurity potential matrix:
  
\begin{equation}
\label{eq23}
{\mb V}_l  = \left(\begin{array}{cc}
\varepsilon_l & 0 \\
0 & -\varepsilon_l
\end{array}
\right).
\end{equation}

To the lowest order in $\varepsilon_i$ we get:
\begin{equation}
\label{eq24}
\mb G(i,j;\omega) = \mb G^0(i,j;\omega)+\sum_n \mb G^0(i,n;\omega)\mb
V_n\mb
G^0(n,j;\omega). 
\end{equation}

Following Eqs. (\ref{eq3}) we  express the interesting
quantities $\Delta_{ij}$ and $\Delta_{il}$, where  $l$ and $j$ are
nearest 
neighbours of $i$, in the lowest
order
of  $\varepsilon_i$ perturbations by means of the disordered Green
function
(Eq. \ref{eq24}) and calculate the mean square deviation  of the pairing
parameter as:
\begin{equation}
\label{eq25}
 <\delta \Delta_{ij} \delta \Delta_{il}^*>=
<\Delta_{ij} \Delta_{il}^*>-<\Delta_{ij}> <\Delta_{il}^*>,
\end{equation}
where bonds $ij$ and $il$ in the pairing potentials  $\Delta_{ij}$ and
$\Delta_{il}$, denote bonds which can be chosen as parallel 
(then $j$ and $l$ coincide) as well as  perpendicular.

The assumption is that random site energies $\varepsilon_i$ in Eqs.
(\ref{eq1},\ref{eq22},\ref{eq23},\ref{eq24}) are
independent variables; then averages
$<\varepsilon_i>$ ,$<\varepsilon_i>$ being independent of the site
index $i$ and $<\varepsilon_i \varepsilon_j> = < \varepsilon_i^2 > \delta_{ij}$
lead to 
\begin{equation}
< \delta \Delta_{ij} \delta \Delta_{il}^*>
= \Gamma_{ij}^l < \varepsilon_i^2 >,
\label{eq26}  
\end{equation} 
where $l=j$ for parallel ($\Gamma_{ij}^j=\Gamma_{ij}^{\parallel}$) and $l=j+1$ perpendicular bonds
($\Gamma_{ij}^{j+1}=\Gamma_{ij}^{\perp}$),
respectively (Fig. 4).
Thus, fluctuations are determined by the corresponding coefficients
$\Gamma^{\parallel}_{ij}$ and
$\Gamma^{\perp}_{ij}$. 
They can be calculated from Eqs. (\ref{eq3}) and (\ref{eq22}-\ref{eq26}): 
\begin{eqnarray}
\Gamma^{l}_{ij} &=& \frac{W_{il} W_{ij}}{\pi^2}
  \frac{1}{N} \sum_{\mb q}
\left( \frac{1}{N} \sum_{\mb k} \int_{-\infty}^{\infty}~{\rm d} \omega~ \frac{{\rm Im} \left(\mb
G^0 (\mb k;
 \omega) \mb
\tau_3 \mb
G^0 (\mb k - \mb q; \omega ) \right)_{12}}
{{\rm e}^{\beta \omega}+1}~ \right. \nonumber \\ &\times& \left. {\rm e}^
{-{\rm i} \mb k
\left(\mb R_i
-\mb R_l \right)} \right)  \left(\frac{1}{N} \sum_{\mb k'} \int_{-\infty}^{\infty}~{\rm d} \omega'~
\frac{{\rm Im} \left(\mb G^0(\mb k'; \omega') \mb \tau_3 \mb
G^0(\mb k' - \mb q; \omega' ) \right)_{12}}{{\rm e}^{\beta \omega}+1}~ \right.
\nonumber \\
&\times& \left. {\rm
e}^{{\rm i} \mb k' \left(\mb R_i
-\mb R_j \right)} \right)  {\rm e}^{{\rm i} \mb q \left(\mb R_j -\mb R_l
\right)}.
\label{eq27}
\end{eqnarray}
After  evaluation at $T=0$ K it yields:
\begin{eqnarray}
\Gamma^{l}_{ij} &=& \frac{1}{N} \sum_{\mb q} \left[\frac{W_{ij}}{N
\pi}
\sum_{\mb k}
\int_{-\infty}^0~{\rm d} \omega~{\rm Im} \left\{ G_{11}^0(\mb k; \omega)
G_{12}^0(\mb k-\mb q; \omega) \right. \right. \nonumber \\ 
\label{eq28}
&-& \left.
\left. 
G_{12}^0(\mb k; \omega)
G_{22}^0(\mb k-\mb q; \omega) \right\}   {\rm e}^{-{\rm i} \mb k \left(\mb
R_i -\mb
R_j \right)}\right] \\
&\times& \left[\frac{W_{il}}{N \pi}
\sum_{\mb k'}
\int_{-\infty}^0 {\rm d}~ \omega'~{\rm Im}
 \left\{ G_{11}^0 (\mb k'; \omega')
G_{12}^0(\mb k'-\mb q; \omega') \right. \right. \nonumber \\ &-& \left.
\left. 
G_{12}^0(\mb k'; \omega)
G_{22}^0(\mb k'-\mb q; \omega) \right\}   {\rm e}^{{\rm i} \mb k \left(\mb
R_i -\mb
R_j \right)}\right] {\rm e}^{{\rm i} \mb q \left(\mb R_j -\mb R_l
\right)}. \nonumber
\end{eqnarray}

\begin{figure}[htb]
\hspace{3cm}
\resizebox{0.3\textwidth}{!}{%
  \includegraphics{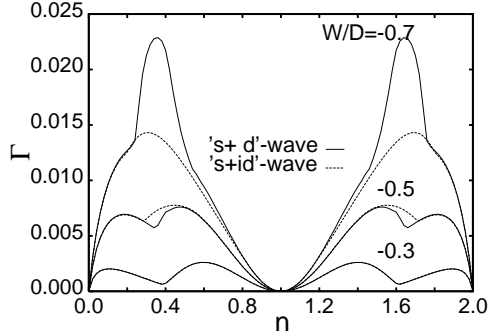}}
\caption{ Fluctuation parameter
$\Gamma=\Gamma^{||}$, versus band filling $n$ for
a pairing potential $\Delta_{ij}$ for three
values of intersite attraction $|W|/D=0.3$, $0.5$, $0.7$.
 Full lines
correspond to
 the real solution '$s+d$' while dashed lines  to
the complex one  '$s+{\rm i} d$' respectively.}
\end{figure}

Depending on site $l$  we get the formula for $\Gamma^{\parallel}_{ij}$ if
sites
$j$ and $l$ are identical   
($ \mb
R_l$  =$\mb R_j$) and $\Gamma^{\perp}_{ij}$ for $l=j+1$ (Fig. 4).
After the integration over $\omega$, and $\omega'$   
(Eq. \ref{eq28}) we get following
formulae:

\begin{eqnarray}
\Gamma^{\perp}_{ij} &=&  \frac{1}{N} \sum_{\mb q}
\left[\frac{W_{ij}}{2N} 
\sum_{\mb k}
\frac{ \sub{\Delta}{k} \sub{\tilde{\epsilon}}{k} + \sub{\Delta}{k} \sub{\tilde{\epsilon}}{
k - q}} {( 
\sub{E}{k}+ \sub{E}{k - q})
 \sub{E}{k} \sub{E}{k - q}}{\rm e}^{{\rm i} \mb k \left(\mb R_i -\mb R_j
\right)
} \right] \nonumber \\
&\times& \left[\frac{W_{il}}{2N}
\sum_{\mb k}
\frac {\sub{\Delta}{k}^* \sub{\tilde{\epsilon}}{k} + \sub{\Delta}{k}^* 
\sub{\tilde{\epsilon}}{ k - q}} {(
\sub{E}{k}+ \sub{E}{k - q})
 \sub{E}{k} \sub{E}{k - q}}{\rm e}^{-{\rm i}  \mb k  \left(\mb R_i -\mb
R_j \right)} 
\right] {\rm e}^{{\rm i} \mb q \left(\mb R_j -\mb R_l \right)}
\label{eq29}
\\
\Gamma^{\parallel}_{ij} &=&  \frac{1}{N} \sum_{\mb q}
\left|\frac{W_{ij}}{2N}
\sum_{\mb k}
\frac {\sub{\Delta}{k} \sub{\tilde{\epsilon}}{ k} + \sub{\Delta}{k} 
\sub{\tilde{\epsilon}}{k - q}} {(
\sub{E}{k}+ \sub{E}{k - q})
 \sub{E}{k} \sub{E}{k - q}}{\rm e}^{\mb k (\mb R_i -\mb R_j)}
\right|^2. \nonumber
\end{eqnarray}

In Fig. 6 we plot the coefficient  $\Gamma^{\parallel}$ versus band filling
$n$.      
Full and dashed lines correspond to '$s+d$' and                          
'$s+{\rm i} d$' solutions, respectively. 
One can see that in case of weak interactions the results are very
similar,
while for large interaction ($W/D=-0.7$) where mixed solutions are present
($n \approx 0.4$)
we observe a large difference. Clearly, in case of a purely real solution 
the fluctuations are larger. Interestingly, we find $\Gamma^{\parallel}=0$
for half filled band $n=1$. This result corresponds to the similar one 
 in case of attractive 'on site' (negative $U$) interaction
\cite{Gyo97,Lit98b,Mor01,Lit96}
and this is due to the particle hole symmetry.

\begin{figure}[htb]
\hspace{-1cm}
\resizebox{0.3\textwidth}{!}{%
  \includegraphics{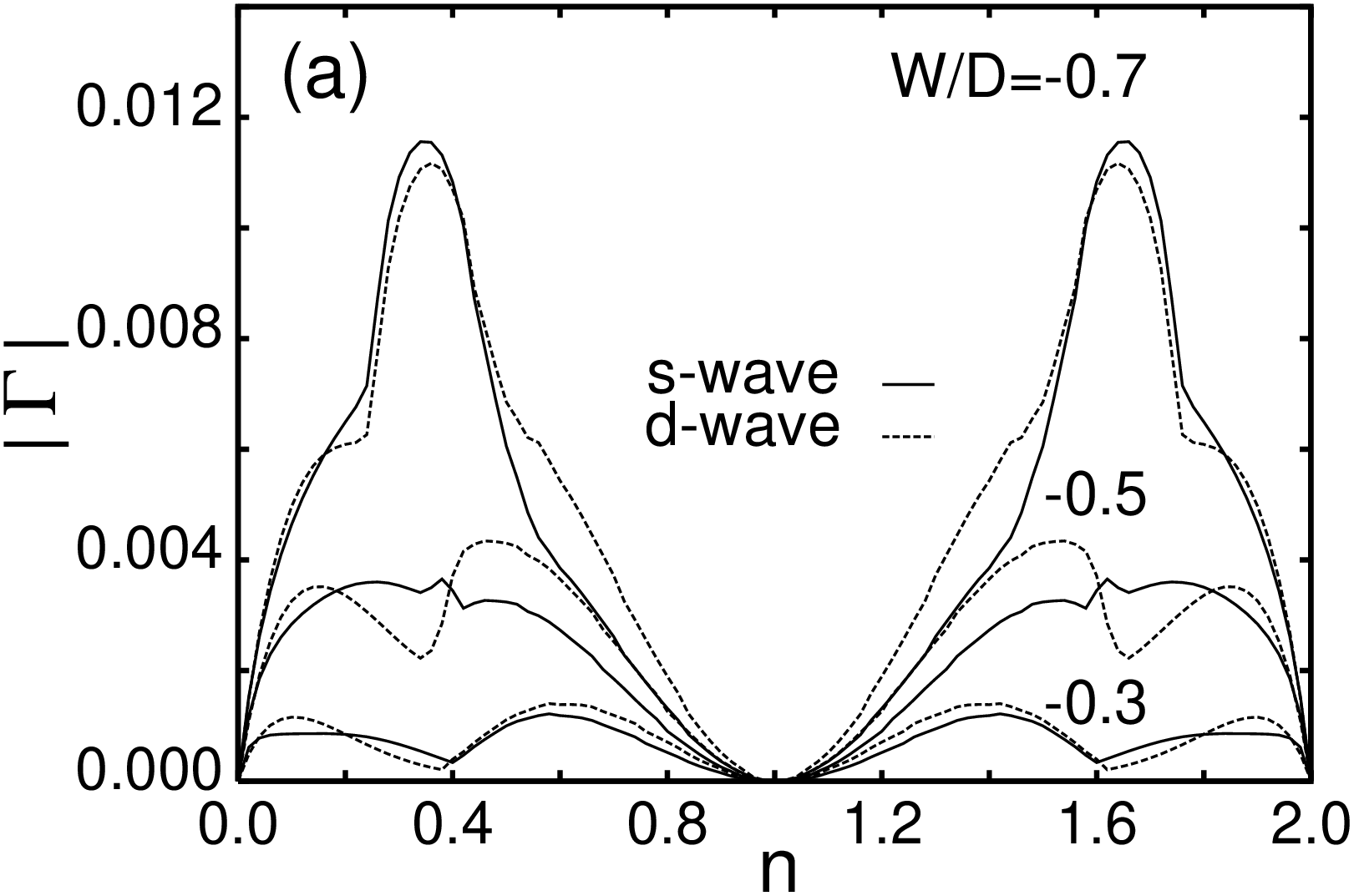}}
\hspace{2.5cm}
\resizebox{0.3\textwidth}{!}{%
  \includegraphics{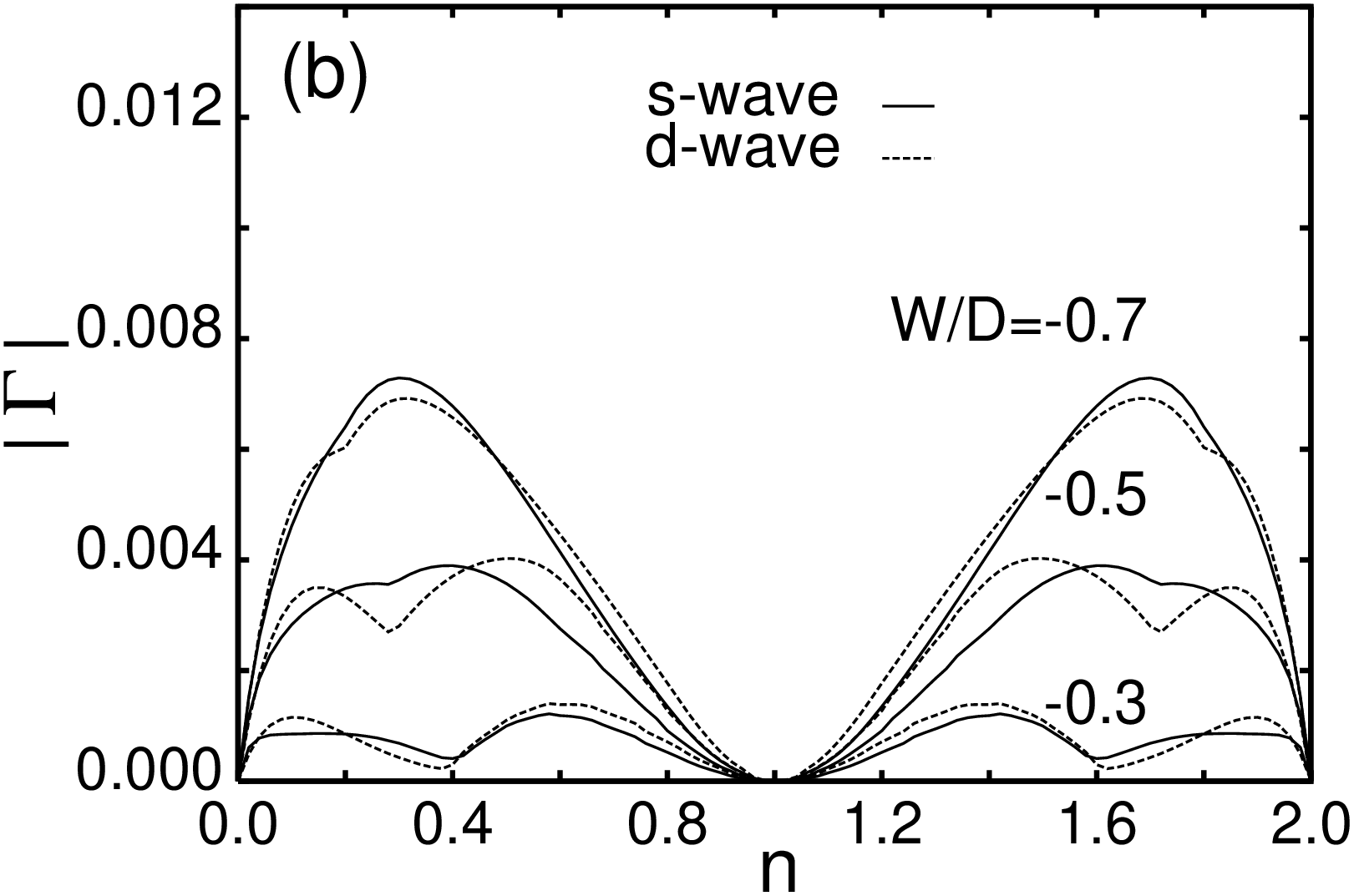}}
\caption{
 Fluctuation parameter
$\Gamma=\Gamma^s$, $\Gamma^d$ versus band filling $n$ for
 $s$- and $d$-
wave pairing potential
(full  and dashed lines respectively)
(dashed
and full lines respectively) for three
values of intersite attraction $|W|/D=0.3$, $0.5$, $0.7$.
Figure (a) corresponds to the real solution '$s+d$' while
(b) to the complex one  '$s+{\rm i} d$' respectively.}
\end{figure}
It is also possible to analyze fluctuations of each of the components of 
the superconducting order parameter (Eq. \ref{eq17}).
 The corresponding coefficients $\Gamma_{ij}^s$ and $\Gamma_{ij}^d$
determine
   fluctuations of the pairing parameter of $s$--wave symmetry $\Delta^s_{ij}$
   and $d$--wave one $\Delta^d_{ij}$.
Namely:
\begin{eqnarray}
< \delta \Delta_{ij}^{s} \delta \Delta_{ij}^s>
= \Gamma^{s}_{ij} < \varepsilon_i^2 > \nonumber \\
< \delta \Delta_{ij}^{d} \delta \Delta_{ij}^d>
= \Gamma^{d}_{ij} < \varepsilon_i^2 >
\label{eq30}
\end{eqnarray}
where we have defined
\begin{eqnarray}
\label{eq31}
\Gamma_{ij}^s &=& \frac{1}{2} (\Gamma_{ij}^{\parallel} +\Gamma_{ij}^{\perp}) \\
\Gamma_{ij}^d &=& \frac{1}{2} (\Gamma_{ij}^{\parallel} -\Gamma_{ij}^{\perp}) \nonumber
\end{eqnarray}

The parameters $\Gamma^{s(d)}$ governing 
 disorder induced fluctuations of the order parameter $\Delta_s(d)$ are
plotted versus band filling in Fig. 7. Figure 7a corresponds to the 
real combination of 
'$s+d$' solution 
while
Fig. 6b corresponds to the complex one  '$s+{\rm i} d$'.  
Note that generally, fluctuations of $'s +d' $ and $d + {\rm i} d$
components are relatively
small $|\Gamma| < 0.012$:
$\Gamma^s$ and $\Gamma^d$ 
are of the same 
order. However their relative value $|\Gamma^s|/|\Gamma^d|$ depends on 
band filling $n$, showing that
the
change of symmetry is possible by disorder. In 
case of particle-hole symmetry at half 
filling both $\Gamma^s$ and
$\Gamma^d$ go to zero (like $\Gamma^{\parallel}$ in Fig. 6).

\section{Conclusions and Discussion}
Solving the extended Hubbard
model (\ref{eq1}), for the appropriate system parameters, we have analyzed possible
singlet solutions with $s$-- and $d$--wave order parameter. Especially we have 
concentrated on
the existence of mixed,   
'$s+d$' and '$s+{\rm i} d$', solutions.
On account of that  the Anderson theorem could not be applied 
to anisotropic superconductors we analyzed the disorder induced, 
spatial
fluctuations of
order parameter amplitudes $\Delta^s$ and $\Delta^d$. Our results
show that the disorder does not induce large fluctuations in the
pairing potentials and the amplitude of pairing potential 
$|\Delta(ij)|$ may rather be
the same for all
bonds in $x$  and $y$ directions.  However, in regions  of mixed solution
the fluctuations
are larger for the '$s+d$' solution, which implies that  '$s+{\rm i} d$'
is
more
stable in the
presence of disorder. Interestingly, we observed zero fluctuations limit in 
case of
particle-hole symmetry as in the case of the negative $U$ Hubbard model
\cite{Gyo97,Lit98b,Mor01,Lit96}.  
This  result  is obtained here
at zero temperature ($T=0$ K)  in the limit
of very weak disorder.
Nevertheless, for  stronger  non-magnetic disorder we observe a pair
breaking effect strongly influencing the critical temperature
\cite{Lit98a,Mar99}. 
Moreover the  effect is stronger if the chemical potential passes 
a Van Hove singularity. 
In our case it is a half--filled situation $n=1$.

Recently Ghosal {\it et al}. \cite{Gho01} have
done
self--consistent calculations
of the pairing amplitude for finite cluster size and various strength
of 
disorder.
Their results, obtained away from half filling $n \neq 1$,
show that in the case of strong enough disorder, large fluctuations of
$\Delta_{ij}$ can lead to
superconducting islands. In that situation the disorder can lead to  phase
fluctuations and to pseudogap phenomena \cite{Tim99}. To investigate
these effects we need to use more sophisticated methods i.e. to go beyond
the Hartree--Fock--Gorkov approximation as in   
Ref. \cite{Gyo91}.

\section*{Acknowledgements}
The author would
like to thank  Prof. K.I. Wysoki\'nski and 
Prof. B.L. Gy\"{o}rffy for discussions.


\begin{thebibliography}{99}
\bibitem{Ann96} J.F. Annett, N. Goldenfeld, and A.J. Legett in {\em Physical Properties
of High temperature Superconductors}, edited by D.M. Ginsberg (World Scientific,
Singapore, 1996), Vol. 5.
\bibitem{Mae01} Y. Maeno, T.M. Rice and M. Sigrist, Physics 
Today 42, Jan. 2001.
\bibitem{And59} P.W. Anderson,   
 J. Phys. Chem. Solids {\bf 11} (1959) 26.
\bibitem{Abr59} A.A. Abrikosov and L.P. Gorkov, Sov. Phys. JETP
{\bf 8}, 1090 (1959).
\bibitem{Mak69} K. Maki in {\em Superconductivity}, edited by R.D. Parks (Marcel
Dekker, New York 1969)
Vol.
2, Chapter 8.
\bibitem{Lus73} H. Lustfeld, J. Low. Temp. Physics. {\bf 12}, (1973) 595.  
\bibitem{Gor83} L.P. Gorkov and P.A. Kalugin JETP Lett. {\bf 41} (1983) 253.
\bibitem{Bor94} L.S. Borkowski and P.J. Hirschfeld, Phys. Rev. B {\bf 49}
(1994) 15404.
\bibitem{Gyo97} B.L. Gy\"{o}rffy, G. Litak and K.I. Wysoki\'nski, in
{\em Fluctuation Phenomena in High Temperature Superconductors},
edited by M. Ausloos and A.A. Varlamov (Kluver, Dordrecht 1997) p. 385.
\bibitem{Lit98b} G. Litak, B.L. Gy\"{o}rffy, K.I. Wysoki\'nski
Physica {\bf C 308} (1998) 132.
\bibitem{Lit98a}  G. Litak, A.M. Martin, B.L. Gy\"{o}rffy,
J.F. Annett and K.I. Wysoki\'nski, Physica C {\bf 309} (1998) 257.
\bibitem{Mar99} A.M. Martin, G. Litak, B.L. Gy\"{o}rffy, J.F. Annett and
 K.I. Wysoki\'nski, Phys. Rev. B {\bf 60} (1999) 7523.
\bibitem{Lit00}  G. Litak, J.F. Annett, B.L. Gy\"{o}rffy,  Acta
Phys. Pol.  A {\bf 97} (2000) 249.
\bibitem{Bon93} W.N. Hardy, D.A. Bonn, D.C. Morgan, R.X. Laing and
K. Zhang, Phys. Rev. Lett. {\bf 70},
(1993) 3999.
\bibitem{Bon94} D.A. Bonn, S. Kammal, K. Zhang, R.X. Liang, D.J. Baar,
E. Kleinand W.N. Hardy, Phys. Rev. B {\bf 50}, (1994) 4051.
\bibitem{Kar00} K. Karpi\'nska, M.Z. Cieplak, S. Guha A. Malinowski, T.
Sko\'skiewicz, W. Plesiewicz, M. Berkowski, B. Boyce, TR Lemberger and P.
Lidenfeld, Phys. Rev. Lett. {\bf 84}, (2000) 155.
\bibitem{Ber96} C. Bernhard, J.L. Tallon, C. Bucci, R. De Renzi, G. Guidi,
G.V.M. Williams, and Ch. Niedermayer, Phys. Rev. Lett. {\bf  77}, 2304
(1996) 2304.
\bibitem{Tal97} J. L. Tallon, C.  Bernhard, G.V.M. Williams, and J.W. Loram,
Phys. Rev. Lett. {\bf 79} (1997) 5294.
\bibitem{Mac98} R.P. Macenzie, R.K.W. Haselwimmer, A.W. Tyler, G.G.   
Lonzarich, 
Y. Mori, S. Nishizaki and Y. Maeno, Phys. Rev. Lett. {\bf 80}, (1998) 161.
\bibitem{Mic90} R. Micnas, J. Ranninger, S. Robaszkiewicz,  Rev.
Mod. Phys. {\bf 62}, (1990) 113.
\bibitem{Mor01} R. Moradian J.F. Annett, B.L. Gy\"{o}rffy, and G. Litak,
Phys. Rev. B {\bf
6302} (2001) 4502.
\bibitem{Lit96} G. Litak, B.L. Gy\"{o}rffy, K.I. Wysoki\'nski, Mol. Phys.
Rep.
{\bf 15-16} (1996) 87.
\bibitem{Gho01} A. Ghosal, M. Randeria, and N. Trivedi,
 Phys. Rev. B {\bf 6302} (2001) 505. 
\bibitem{Tim99} T. Timusk, B. Stat, Rep. Prog. Phys. {\bf 62} (1999) 61. 
\bibitem{Gyo91} B.L. Gy\"{o}rffy, J.B. Stauton, G.M. Stocks,
Phys. Rev. B {\bf 44} (1991) 5190.
\end{thebibliography}
\end{document}